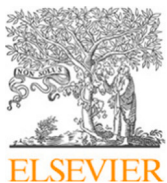

Contents lists available at ScienceDirect

# Medical Image Analysis

journal homepage: www.elsevier.com/locate/media

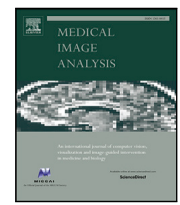

# Uncertainty aware training to improve deep learning model calibration for classification of cardiac MR images

Tareen Dawood [a,*], Chen Chen [c], Baldeep S. Sidhu [a,b], Bram Ruijsink [a,b], Justin Gould [a,b], Bradley Porter [a,b], Mark K. Elliott [a,b], Vishal Mehta [a,b], Christopher A. Rinaldi [a,b], Esther Puyol-Antón [a], Reza Razavi [a], Andrew P. King [a]

[a] School of Biomedical Engineering & Imaging Sciences, King's College London, UK
[b] Guy's and St Thomas' Hospital, London, UK
[c] BioMedIA Group, Department of Computing, Imperial College London, UK

## ARTICLE INFO

## ABSTRACT

Quantifying uncertainty of predictions has been identified as one way to develop more trustworthy artificial intelligence (AI) models beyond conventional reporting of performance metrics. When considering their role in a clinical decision support setting, AI classification models should ideally avoid confident wrong predictions and maximise the confidence of correct predictions. Models that do this are said to be well *calibrated* with regard to confidence. However, relatively little attention has been paid to how to improve calibration when training these models, i.e. to make the training strategy *uncertainty-aware*. In this work we: (i) evaluate three novel uncertainty-aware training strategies with regard to a range of accuracy and calibration performance measures, comparing against two state-of-the-art approaches, (ii) quantify the data (aleatoric) and model (epistemic) uncertainty of all models and (iii) evaluate the impact of using a model calibration measure for model selection in uncertainty-aware training, in contrast to the normal accuracy-based measures. We perform our analysis using two different clinical applications: cardiac resynchronisation therapy (CRT) response prediction and coronary artery disease (CAD) diagnosis from cardiac magnetic resonance (CMR) images. The best-performing model in terms of both classification accuracy and the most common calibration measure, expected calibration error (ECE) was the Confidence Weight method, a novel approach that weights the loss of samples to explicitly penalise confident incorrect predictions. The method reduced the ECE by 17% for CRT response prediction and by 22% for CAD diagnosis when compared to a baseline classifier in which no uncertainty-aware strategy was included. In both applications, as well as reducing the ECE there was a slight increase in accuracy from 69% to 70% and 70% to 72% for CRT response prediction and CAD diagnosis respectively. However, our analysis showed a lack of consistency in terms of optimal models when using different calibration measures. This indicates the need for careful consideration of performance metrics when training and selecting models for complex high risk applications in healthcare.

## 1. Introduction

Artificial intelligence (AI) techniques have the potential to be used as decision support tools in medicine, for example in applications such as diagnosing disease or predicting response to treatment. However, in recent years even though AI models have dominated medical research they are often developed without consideration of how the models will be used in clinical practice. Specifically, the lack of trust in automated predictions for clinical applications is a major barrier preventing clinical adoption (Linardatos et al., 2021). One way to provide a level of trust in AI predictions is to estimate uncertainty or classification confidence and provide end-users with the confidence score as well as the prediction.

Ideally, models used in a decision support setting should avoid confident wrong predictions and maximise the confidence of correct predictions. The concept of *model calibration* refers to the relationship between the accuracy of predictions and their confidence: a well-calibrated model will be less confident when making wrong predictions and more confident when making correct predictions. With this in mind measures of model calibration have been proposed that provide a more complete understanding of the performance of a predictive model by estimating how closely the predictive confidence matches its accuracy (Nixon et al., 2019). However, relatively little attention has been paid to how to optimise AI models with respect to model calibration. If training could be performed in such a way as to maximise both accuracy and calibration this would have the potential to provide a level of trust and reliability in model outputs (Gawlikowski et al., 2021; Sensoy et al., 2021).

In this paper we investigate training schemes that aim to improve model calibration as well as accuracy, with a specific focus on deep learning (DL) models. These schemes are collectively known as *uncertainty-aware* training methods. We utilise recent advances in uncertainty estimation and uncertainty-aware training to investigate multiple methodologies to identify the best performing strategy with respect to accuracy and calibration. Specifically, we investigate two applications from cardiology: prediction of response to cardiac resynchronisation therapy (CRT) from pre-treatment cardiac magnetic resonance (CMR) images, and diagnosis of coronary artery disease (CAD), again from CMR images. In this introduction, we first focus on techniques utilised to estimate uncertainty, followed by a discussion on model calibration and then we move on to methods to develop uncertainty-aware AI models. Finally, we provide an overview of the contributions of our research.

### 1.1. Uncertainty estimation

Two commonly identified sources of uncertainty are aleatoric uncertainty, which is caused by noisy data inputs and epistemic uncertainty, which is the uncertainty inherent in the model itself (Hüllermeier and Waegeman, 2021). Aleatoric uncertainty is irreducible as the 'noise' present in the input data cannot be altered. Epistemic uncertainty, however, may be improved by providing more knowledge through larger and more varied datasets (Abdar et al., 2021; Gawlikowski et al., 2021).

Epistemic and aleatoric uncertainty estimates for task DL models have predominantly been made using Bayesian approximation, ensemble methods and test-time augmentation (Abdar et al., 2021; Gawlikowski et al., 2021). Bayesian DL aims to model a distribution over the model's weights and is a favoured method for uncertainty estimation, as the modelling of an approximated posterior distribution provides the ability to produce more representative epistemic uncertainty estimations (Abdar et al., 2021). However, approximation methods are required to compute the estimates requiring more computational effort for both training and inference (Gawlikowski et al., 2021; Alizadehsani et al., 2021). Ensemble methods seek to train multiple models, each with different parameters, which are then used to generate multiple predictions from which the variance in predicted classes can be considered a measure of the epistemic uncertainty (Gawlikowski et al., 2021). For example, Mehrtash et al. (2020) demonstrated the use of ensembles to quantify a model's predictive uncertainty for medical image segmentation, using MR images of the brain, heart and prostate. Aleatoric uncertainty is often estimated by augmenting test data to generate multiple test samples and measuring the variance in the predictions whilst keeping the model architecture intact (Shorten and Khoshgoftaar, 2019). An example of this type of approach is Wang et al. (2018), who investigated test-time uncertainty estimation to improve automatic brain tumour segmentation tasks using random flipping and rotation, later expanding their research to epistemic uncertainty (Wang et al., 2019).

Currently these approaches have predominantly been applied to medical image segmentation applications and less so for classification applications such as predicting diagnosis or treatment response (Abdar et al., 2021; Gawlikowski et al., 2021). Therefore, actively researching and improving uncertainty estimation techniques to identify more calibrated and easily scaleable estimates will aid the development of trustworthy decision support tools for clinicians (Gawlikowski et al., 2021).

### 1.2. Model calibration

Quantifying uncertainty of DL models has highlighted underlying problems of DL architectures. In particular, the Softmax probability function, often used as the final layer of a DL classification model has been shown to provide over-confident predictions for both in and out of distribution data (Kompa et al., 2021; Gawlikowski et al., 2021). Additionally, the hard label binary classification approach has been shown to have a negative impact by overestimating confidence in predictions, indicating that a softer approach may provide a more reliable method mimicking real world behaviour (Thulasidasan et al., 2019). Guo et al. (2017) highlighted that while developments have been made to produce a variety of architectures and uncertainty estimations for DL models, evaluating the calibration of models is necessary to understand and interpret probability estimates. To this end, Guo et al. (2017) proposed the Expected Calibration Error (ECE) metric, which partitions or bins confidences and utilises the accuracy and confidence estimates over all sets of samples in all bins to provide a measure of model calibration. Interestingly, Nixon et al. (2019) investigated the shortfalls of the ECE, noting that the choice of the number of bins has the potential to skew results. This influence is noticeable when visualised on an illustrative representation of the ECE referred to as a reliability diagram. Responding to this weakness, an alternate measure called the Adaptive ECE (AECE) has been suggested based on an adaptive binning strategy (Ding et al., 2020a). The authors argue that AECE provides a robust approach to handle non-uniform confidence calibration and enables enhanced visual illustrations in reliability diagrams. Often calibration errors are also evaluated with the Maximum Calibration Error (MCE), which quantifies the largest deviation across the confidence bins (Guo et al., 2017). Overconfidence Error (OE) is an additional calibration performance metric which penalises predictions by the weight of the confidence but only when confidence exceeds accuracy (Thulasidasan et al., 2019). OE has been proposed as an appropriate calibration metric for high risk applications such as healthcare where it is important to avoid confident wrong predictions (Thulasidasan et al., 2019). Alternate metrics such as the Brier Score (BS) have been utilised in the literature and considered as a proper scoring rule, computed using uncertainty, resolution and reliability. However, the measure has the potential to under-penalise predictions with lower probabilities.

Despite the range of metrics presented, studies continue to investigate alternate, standardised and improved methods to understand and evaluate the calibration of DL models (Ashukha et al., 2020; Ovadia et al., 2019). To date, most of this research has focused on computer vision problems, and little work has evaluated the utility of these measures on real-world medical applications.

### 1.3. Uncertainty-aware training

Uncertainty-aware training refers to methods that incorporate uncertainty information into the training of a DL model with the aim of improving its model calibration. Yu et al. (2019) provide an example of this type of approach, demonstrating how a DL model can learn to gradually exploit uncertainty information to provide more reliable predictions for a 3D left atrium segmentation application. Alternate approaches aim to directly target confident cases based on an acceptable user-risk level, for example the 'selective' image classification method proposed in Geifman and El-Yaniv (2017). Uncertainty estimates have been directly incorporated into the loss function of the model, as proposed by Ding et al. (2020b) for a segmentation task. The outcomes demonstrated the ability to maximise performance on confident outputs and reduce overconfident wrong predictions. In our previous work (Dawood et al., 2021), we used a similar approach to Ding et al. (2020b) and proposed, for the first time, an uncertainty-aware DL model for CRT response prediction, as a preliminary investigation to evaluate changes in predictive confidence. We used confidence bands estimated at test time to highlight an improvement in the confidence of correct predictions and a reduction in confidence of incorrect predictions. Another group of methods has attempted to define differentiable loss terms that directly quantify model calibration (Krishnan and Tickoo, 2020; Karandikar et al., 2021). Alternately, building on recent work on Evidential DL (Sensoy et al., 2018, 2020), a Bayesian methodology was incorporated into Evidential DL by Sensoy et al. (2021). They utilised probability distributions to obtain uncertainty in predictions for each category/class and introduced new methods to handle the risk associated with incorrect predictions. The continued research and improvements within the field therefore highlight the need to incorporate uncertainty estimation when training DL models as it will likely become a vital component in high-risk applications such as diagnostic predictions in healthcare (Gawlikowski et al., 2021).

### 1.4. Contributions

In this paper, we seek to perform a thorough investigation of uncertainty-aware DL training methods and evaluate them on two real-world clinical applications. Our contributions are:

1. We propose three novel uncertainty-aware training strategies (including the one proposed in our preliminary work (Dawood et al., 2021)), and compare them to two state-of-the-art methods from the literature.
2. We evaluate all models on two realistic medical imaging applications: CRT response prediction and CAD diagnosis, both from CMR images.
3. We use a wide range of calibration performance measures proposed in the literature, combined with a reliability diagram based on adaptive binning to understand the effects of different uncertainty-aware training methods.
4. We further quantify the performance of all models in terms of aleatoric and epistemic uncertainty.
5. We evaluate the impact of using a calibration-based model selection criterion on accuracy and calibration performance.

The paper is structured as follows. In Section 2 we describe our uncertainty-aware strategies and the comparative approaches. In Section 4 we present all experiments performed to evaluate and compare the different approaches with all results found in Section 5. Section 6 then discusses the findings, evaluates the outcomes and recommends future work towards cultivating trustworthy and calibrated predictive DL classification models.

## 2. Methods

In this Section we introduce the different uncertainty-aware and comparative strategies used and evaluated in the paper.

### 2.1. Notation

Before presenting our novel and comparative approaches for uncertainty-aware training, we first define a common notation that will be used in the subsequent descriptions. Throughout, $\mathbb{P}$ represents the probability of an event, $A \cap B$ represents the intersection of set $A$ and set $B$, $A \cup B$ represents their union, $|A|$ denotes the cardinal of the set $A$ and $\bar{A}$ its complement. Hyperparameters of the network are denoted by Greek lowercase letters: $\theta$ is trainable whereas $\lambda$, $w$ and $\mu$ are hyperparameters. We denote by $B$ the set of samples in a training batch. For a binary classifier $f_\theta$ with trainable parameters $\theta$, we define:

- $G \subset B$ the samples labelled as ground truth positive ($G \cup \bar{G} = B$)
- $P_\theta \subset B$ the samples classified by the model as positive ($P_\theta \cup \bar{P}_\theta = B$)
- $\mathcal{T}_\theta = (P_\theta \cap G) \cup (\bar{P}_\theta \cap \bar{G})$ the samples correctly classified ($\mathcal{T}_\theta \cup \bar{\mathcal{T}}_\theta = B$)

- $\mathcal{U}_\theta \subset B$ the samples with an "uncertain" classification (based on their classification confidence) ($\mathcal{U}_\theta \cup \bar{\mathcal{U}}_\theta = B$)
- $f_\theta : x \to [\mathbb{P}(x \in P_\theta), \mathbb{P}(x \in \bar{P}_\theta)]$ (for a sample $x$ from $B$)
- $r_i$ is the confidence (probability) of the model-predicted class for sample $i$, i.e. $r_i = max[\mathbb{P}(x_i \in P_\theta), \mathbb{P}(x_i \in \bar{P}_\theta)]$
- $g_i$ is the ground truth label of sample $i$ (i.e. 1 for positive and 0 for negative)

### 2.2. Baseline model

The diagram in Fig. 1 illustrates the architecture of the baseline classification model developed by Puyol-Antón et al. (2020), Dawood et al. (2021), which was used as the framework to perform the experiments. The baseline model utilises CMR short axis (SA) image segmentations produced by a pre-trained U-net (Chen et al., 2020). These segmentations are used as input into a variational autoencoder (VAE), which during the training phase is tasked with reconstructing the segmentations frame-by-frame from the learned latent representations. Subsequently, a classifier is trained to make predictions from the concatenated VAE latent spaces of the time series of CMR SA segmentations.

The points at which aleatoric uncertainty and epistemic uncertainty are estimated (see Section 5.3) are shown in the dotted blocks in Fig. 1. In the literature, quantifying aleatoric uncertainty has often been performed using data augmentation at test time (Ayhan and Berens, 2018). In our work, we produced realistic augmentations for this purpose by using the U-net segmentation model to generate multiple segmentations which were inputted into the VAE/classifier to estimate aleatoric uncertainty. To quantify epistemic uncertainty we drew multiple samples from the learned VAE latent space.

Formally, we define the loss function of the baseline model as comprising three terms:

$$\mathcal{L} = \mathcal{L}_{RE} + \lambda_{KL}\mathcal{L}_{KL} + \lambda_C\mathcal{L}_C \tag{1}$$

In the above equation, $\mathcal{L}_{RE}$ is the cross entropy between the input segmentations to the VAE and the output reconstructions, $\mathcal{L}_{KL}$ is the Kullback–Leibler (KL) divergence between the latent variables and a unit Gaussian, $\mathcal{L}_C$ represents the binary cross entropy loss for the classification task and $\lambda_{KL}$, $\lambda_C$ are used to weight the level of influence each term has to the total loss.

### 2.3. Novel approaches

We now propose three novel approaches to uncertainty-aware training, each based on modifying the baseline model with the aim of improving model calibration. Two weighted loss terms are introduced in Sections 2.3.1 and 2.3.2 respectively. For these a new loss term $\mathcal{L}_N$ and hyperparameter $\lambda_N$ are added to the baseline Eq. (1) to define the general form below:

$$\mathcal{L} = \mathcal{L}_{RE} + \lambda_{KL}\mathcal{L}_{KL} + \lambda_C\mathcal{L}_C + \lambda_N\mathcal{L}_N \tag{2}$$

Furthermore, in Section 2.3.3 we develop a confidence-based weighting scheme applied to the classifier loss $\mathcal{L}_C$.

### 2.3.1. Paired Confidence Loss

We introduced this method in our preliminary work in Dawood et al. (2021). This approach was inspired by the work of Ding et al. (2020b), who proposed an uncertainty-aware training method for image segmentation that focused the loss function on more confident outputs. We adapted this approach to image-based classification problems. The final loss function implemented is:

$$\mathcal{L}_U(\theta) = \sum_{S_\theta \in \{P_\theta, \bar{P}_\theta\}} \frac{1}{|\bar{\mathcal{T}}_\theta \cap S_\theta|} \sum_{(x_i, x_j) \in \bar{\mathcal{T}}_\theta \cap S_\theta \times \mathcal{T}_\theta \cap S_\theta} max(\mathbb{P}(x_i \in S_\theta) - \mathbb{P}(x_j \in S_\theta) + \mu, 0) \tag{3}$$

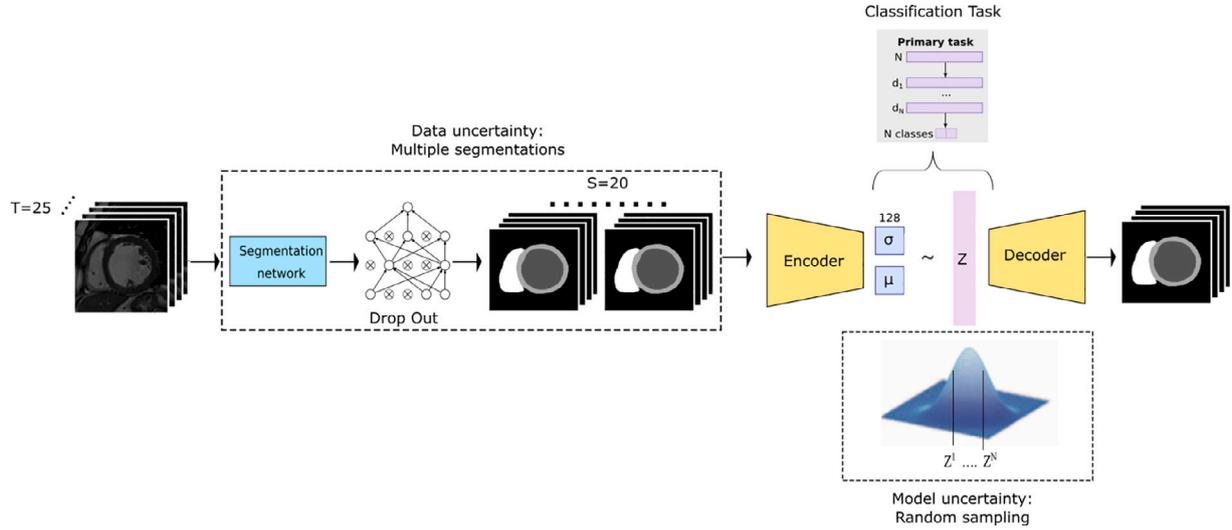

**Fig. 1.** Diagram showing the architecture of the baseline VAE/classification model developed by Puyol-Antón et al. (2020), Dawood et al. (2021) together with the stages at which uncertainty estimates are made. The CMR SA images are segmented using a DL-based model, and these segmentations act as the inputs to the VAE. The classification is performed in the latent space of the VAE and the points at which aleatoric uncertainty and epistemic uncertainty were estimated are shown using dotted blocks.

Here, the first sum is over samples ($S_\theta$) classified as positive ($P_\theta$) or negative ($\bar{P}_\theta$) in a batch, and in the second sum the ($x_i$, $x_j$) are pairs of false positive (or negative) and true positive (or negative) samples.

Intuitively, Eq. (3) will evaluate all pairs of correct/incorrect positive/negative predictions in a training batch, and the terms will be positive when the incorrect prediction ($i$) has higher confidence than the correct one ($j$). If the correct one has higher confidence than the incorrect one by a margin of the hyperparameter $\mu$ or more the terms will be zero. Note that in the *max* term of Eq. (3), the probability of a *correct* prediction is subtracted from the probability of an *incorrect* prediction.

### 2.3.2. Probability Loss

In our second novel method, we again adapted the baseline model loss function to more heavily penalise incorrect predictions with high confidence. We note that the standard cross entropy loss already penalises such cases. However, it is well-known that models trained with cross-entropy loss are prone to poor calibration (Guo et al., 2017), and this motivated the formulation of the Probability Loss approach:

$$\mathcal{L}_{\mathrm{p}}(\theta) = \frac{1}{|G|} \sum_{x_i \in G} \mathbb{P}(x_i \in \bar{P}_\theta) + \frac{1}{|\bar{G}|} \sum_{x_i \in \bar{G}} \mathbb{P}(x_i \in P_\theta) \quad (4)$$

As before, the developed loss term is added into the loss function of the model to follow the form in Eq. (2). The Probability Loss function differs from the approach described in Section 2.3.1 as the $\mathbb{P}$ terms represent the class probabilities of the classifier (after the Softmax layer) for positive and negative *ground truth* samples. Intuitively, this loss term penalises ground truth positive (negative) samples with high confidence in negative (positive) prediction. The terms are normalised by the number of samples for the positive and negative classes in the training batch.

### 2.3.3. Confidence Weight

An alternative solution to defining a new loss term is to add a weighting term to the existing classifier loss $\mathcal{L}_C$ to penalise training samples with highly confident incorrect predictions. The weighting term is determined by first estimating the epistemic uncertainty of each prediction in the batch by sampling in the latent space of the VAE. Specifically, we randomly sampled 20 points from the VAE latent space and computed predictions for each one.

The prediction confidence was calculated as the proportion of positive predictions from these samples and we denote this by $C_i \in [0, 1]$ for sample $i$. See Section 5.3 for further details of the epistemic uncertainty estimation. The weighting term for each sample in the batch was computed as follows:

$$\mathcal{W}_i = g_i \cdot (1 - C_i) + (1 - g_i) \cdot C_i \quad (5)$$

Here, $\cdot$ denotes scalar multiplication. Intuitively, these weights will be high when making a confident wrong prediction, thus encouraging the model training to focus on minimising such cases. $\mathcal{W}_i$ was then scaled to produce $\mathcal{W}_{S,i}$ to ensure that the weights would not drop below a pre-defined value $w$:

$$\mathcal{W}_{S,i} = (1 - w) \cdot \mathcal{W}_i + w \quad (6)$$

Note that $w$ is a hyperparameter that is optimised during the training of the classifier.

### 2.4. Comparative approaches

We now present three existing methods proposed in the recent literature as our comparative uncertainty-aware training strategies.

### 2.4.1. Accuracy versus Uncertainty Loss

Recent work by Krishnan and Tickoo (2020) utilised the relationship between accuracy and uncertainty to develop a loss function aimed at improving model calibration. A differentiable Accuracy versus Uncertainty (AvUC) loss function was developed by placing each prediction into one of four categories; accurate and certain, accurate and uncertain, inaccurate and certain and lastly inaccurate and uncertain. Utilising these four categories, a differentiable loss term was defined as follows:

$$\mathcal{L}_{\mathrm{AvUC}}(\theta) = \log \left( 1 + \frac{|\mathcal{T}_\theta \cap \mathcal{U}_\theta| + |\bar{\mathcal{T}}_\theta \cap \bar{\mathcal{U}}_\theta|}{|\mathcal{T}_\theta \cap \bar{\mathcal{U}}_\theta| + |\bar{\mathcal{T}}_\theta \cap \mathcal{U}_\theta|} \right) \quad (7)$$

Similar to the methods proposed in Sections 2.3.1 and 2.3.2, the final loss function follows the same structure as the baseline model and the AvUC loss is added to the total loss, weighted by the hyperparameter $\lambda_N$ as presented in Eq. (2).

### 2.4.2. Soft ECE loss function

Karandikar et al. (2021) extended research performed by Krishnan and Tickoo (2020), leveraging on the approach of a differentiable loss function to improve calibration. However, here they investigated the ECE measure as a differentiable loss. To implement the loss they introduced a soft binning function scaled with a *soft binning temperature*, $T$ (see Section 4.3). Below we define this loss function using our notation, we refer the reader to Karandikar et al. (2021) for a full explanation using the original notation.

$$\mathcal{L}_{SECE}(\theta, M) = \left( \sum_{m=1}^{M} \left( \frac{|B_j|}{|B|} \cdot \left| A_j - R_j \right|^p \right) \right)^{1/p} \quad (8)$$

Here, we use $B_j$ to denote the set of samples that fall into confidence bin $j$, $M$ represents the number of bins, and $m$ and $j$ represent the confidence bins and are related using the soft binning membership function described in Krishnan and Tickoo (2020). $A_j$ represents the average accuracy within bin $j$ (i.e. the proportion of $B_j$ that are correctly classified) and $R_j$ represents the average confidence within bin $j$ (i.e. the average of $r_i$ within $B_j$). The term $p$ is the order of the soft binning function.

### 2.4.3. Maximum Mean Calibration Error loss function

Kumar et al. (2018) utilised a reproducing kernel Hilbert space (RKHS) approach with a differentiable loss function to improve calibration, which they termed the Maximum Mean Calibration Error (MMCE).

$$
\begin{aligned}
\mathcal{L}_{MMCE}(\theta) = & \sum_{(x_i, x_j) \in \hat{\mathcal{T}}_\theta} \frac{r_i \cdot r_j \cdot k(r_i, r_j)}{(|\mathcal{T}_\theta| - |\mathcal{B}|)^2} \\
& + \sum_{(x_i, x_j) \in \mathcal{T}_\theta} \frac{(1 - r_i) \cdot (1 - r_j) \cdot k(r_i, r_j)}{|\mathcal{B}|^2} \\
& - 2 \sum_{(x_i, x_j) \in \mathcal{T}_\theta \cdot \hat{\mathcal{T}}_\theta} \frac{(1 - r_i) \cdot r_j \cdot k(r_i, r_j)}{(|\mathcal{T}_\theta| - |\mathcal{B}|) \cdot |\mathcal{B}|}
\end{aligned}
\quad (9)
$$

Here, $k$ represents the Hilbert space kernel and all other terms are as defined in Section 2.1. Further detailed derivations and explanations using the original notation are provided in Kumar et al. (2018).

## 3. Materials

We performed two experiments utilising different materials for each. The first experiment focused on response prediction for CRT patients, and the second on diagnosis of CAD. See Table 1 for a summary of the data used in each experiment, which are further described below. Both experiments utilised CMR images as model inputs. In both, we train and evaluate the baseline model featuring a segmentation network followed by a VAE and classifier, and compare this with six different uncertainty-aware versions of the same model.

### 3.1. CRT response prediction model

We used two databases to train and evaluate our baseline and uncertainty-aware CRT response prediction models: (i) CMR SA stacks of 10,000 subjects (a mix of healthy and cardiovascular disease patients) from the UK Biobank (UKBB) dataset (Petersen et al., 2015) and (ii) a database from the clinical imaging system of Guy's and St Thomas' NHS Foundation Trust (GSTFT) consisting of 20 heart failure (HF) patients and 73 CRT patients. The UKBB database was utilised to train the VAE, the HF patients for fine-tuning the segmentation model and the VAE and the CRT patients were used to train and evaluate the VAE and classifier. Further details are provided in Section 4.1.

Details of the UKBB database are provided in Section 3.2. For the GSTFT database, all 73 CRT patients met the conventional criteria for CRT patient selection, chosen using current clinical guidelines based on New York Heart Association classification, left ventricular ejection fraction, QRS duration, the type of bundle branch block and etiology of cardiomyopathy and atrial rhythm (Members et al., 2013). CMR imaging was performed prior to CRT and the CMR multi-slice SA stack was used in this study. The Siemens Aera 1.5T, Siemens Biograph mMR

3T, Philips 1.5T Ingenia and Philips 1.5T and 3T Achieva scanners were used to perform CMR imaging. The typical slice thickness was 8-10 mm, in-plane resolution was between $0.94 \times 0.94$ mm$^2$ and $1.5 \times 1.5$ mm$^2$ and the temporal resolution was 13–31 ms/frame. Using post-CRT echocardiography images (at 6 month follow up), a positive response was defined as a 15% reduction in left ventricular (LV) end-systolic volume. The HF patients had similar CMR imaging details to the CRT patients.

For this experiment, for all datasets the top three slices of the SA stack were employed as the input to the models described in Section 2. Ideally all slices should be utilised but for computational efficiency we only used three slices for prediction. We chose the basal to mid slices as these slices exhibit most myocardial deformation throughout contraction (Jung et al., 2006). All slices were resampled in-plane to a voxel size of $1.25 \times 1.25$ mm, cropped to $80 \times 80$ pixels, and temporally resampled to $T = 25$ time samples as per the same process utilised by Puyol-Antón et al. (2020), before being used for training/evaluation of the models.

### 3.2. CAD diagnosis model

For the CAD diagnosis model all images were extracted from the UKBB. Images were obtained on a 1.5 T MRI scanner (MAGNETOM Aera, Siemens Healthcare, Erlangen, Germany). A typical CMR dataset consists of 10 SA image slices with a matrix size of $208 \times 187$ and a slice thickness of 8 mm, covering both ventricles from the base to the apex. The in-plane image resolution is $1.8 \times 1.8$ mm$^2$, the slice gap is 2 mm, with a repetition time of 2.6 ms and an echo time of 1.10 ms. Each cardiac cycle consists of $T = 50$ frames, with further details on the image acquisition protocol described in Petersen et al. (2015). For the CAD diagnosis experiment, we utilised 16022 UKBB subjects (14384 healthy and 1638 CAD subjects). As coronary occlusions can occur throughout the coronary tree we chose the middle three slices of the SA stack to cover the base, mid and apical portions of the heart for all subjects, similar to Clough et al. (2019). All slices were cropped to $80 \times 80$ pixels and did not require any re-sampling. To follow a similar approach to the CRT experiment, only 25 time frames were utilised for the training of the models.

### 3.3. Ethics

Institutional ethics approval was obtained for use of the clinical data and all patients consented to the research and for the use of their data. All relevant protocols were adhered to in order to retrieve and store the patient data and related images.

## 4. Experiments

Below we describe the details of our two experiments. Please refer to Table 1 for summaries of the data used in each.

### 4.1. Experiment 1 - CRT response prediction

In the first experiment the task was to predict the binary response to CRT (positive/negative) using the pre-treatment CMR data. In order to train the framework for this task the following steps were performed:

- *Fine-tune the pre-trained segmentation model*: The segmentation model (Chen et al., 2020) was pre-trained using UKBB CMR data so to make it robust to the clinical GSTFT data it was fine-tuned using CMR data from the 20 GSTFT HF patients. The fine-tuning was carried out using 300 manually segmented CMR SA slices (multiple slices/time points from the 20 CMR scans).
- *Segment the UKBB and GSTFT CRT CMR data*: The fine-tuned segmentation model was used to automatically segment all frames of the 10,000 UKBB subjects as well as the 73 GSTFT CRT subjects. (Note that this cohort of 10,000 UKBB subjects was separate from the UKBB data used to initially train the segmentation model.)

**Table 1**

Summary of datasets used in training/evaluating the different models for the task of CRT response prediction (left) and for the task of CAD diagnosis (right).

| Method | CRT | | | CAD | | |
|---|---|---|---|---|---|---|
| | Segmentation[a] | VAE | Classifier | Segmentation[a] | VAE | Classifier |
| 1. UKBB 10 000 subjects | | ✓ | | | | |
| 2. UKBB 16022 CAD subjects | | | | ✓ | ✓ | ✓ |
| 3. GSTFT 73 CRT subjects | | ✓(fine-tuned) | ✓ | | | |
| 4. GSTFT 20 HF subjects | ✓fine-tuned | ✓(fine-tuned | | | | |

[a]Segmentation model was pre-trained on a separate set of UKBB data.

**Table 2**

Summary of the optimal hyperparameters achieved for the baseline and each uncertainty-aware strategy for the task of CRT response prediction (left) and for the task of CAD diagnosis (right). We refer the reader to Section 2, where all parameter descriptions are presented.

| Method | CRT | | | | | CAD | | | | |
|---|---|---|---|---|---|---|---|---|---|---|
| | $\lambda_{KL}$ | $\lambda_C$ | $\lambda_N$ | $\mu$ | $w$ | $\lambda_{KL}$ | $\lambda_C$ | $\lambda_N$ | $\mu$ | $w$ |
| 1. Baseline model | 0.001 | 3 | – | – | – | 0.1 | 1.3 | – | – | – |
| 2. Paired Confidence Loss | 0.001 | 1.8 | 1 | 0.6 | – | 0.1 | 1.5 | 0.4 | 0.8 | – |
| 3. Probability Loss | 0.1 | 2 | 0.5 | – | – | 0.1 | 0.6 | 1.2 | – | – |
| 4. Confidence Weight | 0.001 | 2 | – | – | 1 | 0.001 | 1.5 | – | – | 2 |
| 5. Accuracy versus Uncertainty Loss | 0.001 | 2 | 2 | – | – | 0.001 | 2 | 3 | – | – |
| 6. Soft ECE Loss | 0.001 | 1.5 | 1 | – | – | 0.001 | 0.6 | 1.5 | – | – |
| 7. MMCE Loss | 0.001 | 3 | 10 | – | – | 0.001 | 3 | 2 | – | – |

- *Train the VAE*: The VAE was pre-trained using the U-net segmented UKBB data and fine-tuned using the ground truth segmentations of the GSTFT HF data.
- *Train the VAE and classifier together*: We then used the U-net segmented CRT data to train the VAE and CRT classifier for 300 epochs similar to Puyol-Antón et al. (2020). For training each uncertainty-aware method, the fine-tuned VAE model was used, the uncertainty-aware loss function or weighting introduced and then both the VAE and CRT classifier trained for 300 epochs using the U-net segmentations of the 73 CRT patients.

In this experiment, the framework was trained using a faster learning rate for the VAE and a slower rate for the CRT classifier ($10^{-2}$ to ($10^{-8}$), with a batch size of 8. For all approaches, the final model was selected as the one with the highest validation balanced accuracy (BACC) over the classifier training epochs.

Both the CRT baseline and uncertainty-aware models were trained and evaluated using a 5-fold nested cross validation. For each of the 5 outer folds, an inner 2-fold cross validation was performed with grid search hyperparameter optimisation over a range of values. In these inner folds, the set of hyperparameters yielding the highest validation BACC was selected. The optimal hyperparameters were used to train a model (using all training data for that outer fold) and then applied to the held-out (outer) fold. This process was repeated for all outer folds. In this way, hyperparameter optimisation was performed using training data and the model was always applied to completely unseen data. Note also that the CRT data had not been used in pre-training either the segmentation model or the VAE. The hyperparameters optimised using grid search for the CRT response prediction model are presented (on the left) in Table 2. The hidden layer size in the classifier was also optimised as a hyperparameter but all methods found an optimal size of 32.

### 4.2. Experiment 2 - CAD diagnosis

In this experiment the task was to diagnose (positive/negative) CAD from CMR images. A similar training procedure was followed as in Experiment 1, i.e.

- *Segment the UKBB CMR data*: First, the U-net segmentation model (Chen et al., 2020) (pre-trained on the separate UKBB cohort as in Experiment 1) was used to segment the 16,022 UKBB CMR stacks. Note that no fune-tuning was necessary for this experiment as it used only UKBB data.

- *Train the VAE*: The VAE was pre-trained using the segmented UKBB CMR data for 60 epochs.
- *Train the VAE and classifier together*: The classifier was introduced and trained for a further 35 epochs. For training the uncertainty-aware methods, the trained VAE was used and the classifier trained for an additional 35 epochs.

In this experiment the framework was trained using the same learning rate for the VAE and classifier ($10^{-5}$) with a batch size of 25. As for the CRT experiment, the highest validation BACC was used for model selection. For validation a single training/validation/test split of 11535/1282/3205 subjects was employed (i.e. 16022 subjects in total, comprising 14384 healthy and 1638 CAD subjects, as detailed in Section 3.2). The same hyperparameters from the CRT application were optimised for the CAD diagnosis model using grid search. The final hyperparameters are presented (on the right) in Table 2. Similar to the CRT classifier all methods had an optimal hidden layer of size 32 across all strategies.

### 4.3. Additional hyperparameters for comparative approaches

In addition to the hyperparameters in Table 2, the AvUC loss function utilised hyperparameters stated in the paper by Krishnan and Tickoo (2020). Specifically, a warm up strategy was employed, starting with the uncertainty threshold set to 1 and then updated every epoch after the first 3 epochs. The additional parameters for the Soft ECE loss function were the same as those stated in Karandikar et al. (2021). We fixed the number of bins $M$ to keep the search space manageable at a value of 15 and varied the *soft binning temperature* or $T$ value to obtain an optimal outcome at $T = 0.1$ for CRT response prediction and $T = 0.01$ for CAD diagnosis. The parameter $T$ is described in detail in the original paper, Karandikar et al. (2021) and is utilised as a parameter to scale the bins or soften them.

### 4.4. Implementation details

All models were trained on a NVIDIA A6000 48 GB GPU using an Adam optimiser. All data for both experiments was augmented with random flipping and rotations. The code[1] and implementation details is available for download and use.

---

[1] https://github.com/tareend/Uncertainty-Aware-Training—Model-Calibration-for-Classification-of-Cardiac-MR-Images.

**Table 3**

The Sensitivity (SEN), Specificity (SPE) and Balanced Accuracy (BACC) for the baseline classifier and the five uncertainty-aware methods for CAD diagnosis. The best performing CRT classifier for each model was chosen using the best validation BACC achieved over 35 epochs. The best-performing CAD classifier for each model was chosen using the best validation accuracy achieved over 300 epochs. The strategy(s) with the highest BACC is indicated in bold. Those with an asterisk indicate statistical significance when compared to the baseline (McNemar's test, 0.05 significance level) which was found in the CAD experiment.

| Method | CRT | | | CAD | | |
|---|---|---|---|---|---|---|
| | SEN (%) | SPEC (%) | BACC (%) | SEN (%) | SPEC (%) | BACC (%) |
| 1. Baseline model | 73.3 | 64.3 | 68.8 | 75.0 | 65.3 | 70.0 |
| 2. Paired Confidence Loss | 57.8 | 78.6 | 68.2 | 61.9 | 78.3 | 70.1* |
| 3. Probability Loss | 68.9 | 64.3 | 66.6 | 79.3 | 58.1 | 68.7* |
| 4. Confidence Weight | 62.2 | 78.6 | 70.4 | 73.5 | 70.2 | 71.9* |
| 5. AvUC Loss | 68.9 | 71.4 | 70.2 | 66.2 | 68.4 | 67.3* |
| 6. Soft ECE Loss | 75.6 | 57.1 | 66.3 | 58.8 | 79.2 | 69.0* |
| 7. MMCE Loss | 71.1 | 67.9 | 69.5 | 64.3 | 75.4 | 70.0 |

# 5. Experimental results

## 5.1. Evaluation metrics

In our work we present a number of performance metrics to evaluate our uncertainty-aware strategies. First, we utilise the conventional classification performance measures: sensitivity, specificity and BACC (Carrington et al., 2021). Second, we include the ECE value (Guo et al., 2017) as a measure of model calibration. The confidence used when calculating ECE was the predicted probability after the Softmax layer. A set number of confidence bins was chosen and the average accuracy achieved by the model for all samples that fall into each confidence bin was computed. We then calculate the ECE as follows:

$$ECE = \sum_{m=1}^{M} \frac{|\mathcal{B}_m|}{n} \left| acc(\mathcal{B}_m) - conf(\mathcal{B}_m) \right| \quad (10)$$

In Eq. (10), the confidences are grouped into $M = 15$ bins, $\mathcal{B}_m$ is the set of samples whose predictions fall into bin $m$, and $n$ is the total number of samples in all the bins, with corresponding accuracies ($acc$) and confidences ($conf$) (Guo et al., 2017).

Our next calibration measure is the Overconfidence Error (OE), which aims to quantify confident wrong predictions and is computed as follows:

$$OE = \sum_{m=1}^{M} \frac{|\mathcal{B}_m|}{n} \left[ conf(\mathcal{B}_m) \cdot \right.$$
$$\left. max\left( conf(\mathcal{B}_m) - acc(\mathcal{B}_m), 0 \right) \right] \quad (11)$$

Once again the Softmax confidences are grouped into $M = 15$ bins, $\mathcal{B}_m$ is the set of samples whose predictions fall into bin $m$, and $n$ is the total number of samples in all the bins, with corresponding accuracies ($acc$) and confidences ($conf$) (Thulasidasan et al., 2019).

Our third calibration measure is the Maximum Calibration Error (MCE), which is based on the ECE equation but finds the maximum calibration error across the bins (Guo et al., 2017).

$$MCE = \max_{m \in (1, \dots, M)} \left| acc(\mathcal{B}_m) - conf(\mathcal{B}_m) \right|$$

Our final calibration measure is the Brier Score (BS), which is a cost function that evaluates the accuracy of probabilistic predictions, using the prediction probability from the Softmax layer, as presented in Eq. (12).

$$BS = \frac{1}{N} \sum_{i=1}^{N} (p_i - o_i)^2 \quad (12)$$

Here, $N$ represents the number of samples, $p_i$ the probability and $o_i$ represents the ground truth one-hot encoded vector. A low BS indicates a well calibrated model.

## 5.2. Evaluation of uncertainty-aware models

### 5.2.1. Accuracy

The performances in terms of classification accuracy of each uncertainty-aware model on both the CRT response prediction and CAD diagnosis tasks are presented in Table 3. Analysing the results we can see that the Confidence Weight term produced the highest test BACC for both tasks. McNemar's non-parametric test was used to test if the baseline classifier versus each of the uncertainty-aware classifiers had statistically significantly different classification performances at a significance level of 0.05. For the CAD model there was a significant difference across all strategies, indicated with asterisks, but for the CRT response models (which had a smaller test set) the tests indicated no statistically significant differences.

### 5.2.2. Calibration

All calibration measures computed are presented in Table 4 for the CRT response prediction and CAD diagnosis models respectively, with all best performing metrics indicated in bold. The experiments were run three times with different random weight initialisations and the mean and standard deviation of all metrics are shown.

For all metrics, a lower score implies a better calibrated model. The most widely used calibration metric in the literature has been the ECE. The results indicate that the Confidence Weight term reduced the ECE measure the most on both the CRT and CAD predictive models. However, this conclusion is not as clear when considering the other calibration metrics, with all tested models (including the baseline) performing best according to at least one metric for one experiment. However, we note that the results for the CRT experiment might be less reliable due to the smaller test set size.

To visualise the calibration performance of the different models, we present reliability diagrams in Figs. 2 and 3 for our larger cohort of CAD subjects using AECE. The reliability diagram plots accuracy against confidence and a perfectly calibrated model would have a line close to identity. We can see that the Confidence Weight model (Fig. 3d) shows the most improvement across the confidence bands, however improvement is still lacking in the high confidence bands.

## 5.3. Uncertainty quantification

To further understand the effect of uncertainty-aware training on model calibration we now estimate the aleatoric and epistemic uncertainties of our different models. The specific points at which uncertainty was estimated are illustrated in Fig. 1. To estimate the aleatoric uncertainty we generated multiple plausible segmentation inputs to the VAE using inference-time dropout in the segmentation model with probability=0.2, similar to Dawood et al. (2021). Aleatoric uncertainty was then estimated using the prediction of the original data's segmentations and those from 19 additional segmentation sets generated in this way, i.e. the original and 19 additional segmentations were propagated through the VAE and classifier. We note that using dropout

**Table 4**

All calibration metrics (Expected Calibration Error (ECE), Adaptive ECE (AECE), Overconfidence Error (OE), Maximum Calibration Error (MCE) and Brier Score(BS)) computed for the CRT (left) and CAD (right) baseline classifier and the five uncertainty-aware methods. All values are mean ± standard deviation computed over three runs. The lowest and optimal metric across strategies is highlighted in bold. All metrics should ideally move to zero as models become more calibrated as indicated by the ⇓.

| Method | CRT | | | | | CAD | | | | |
|---|---|---|---|---|---|---|---|---|---|---|
| | ECE ⇓ | AECE⇓ | OE ⇓ | MCE ⇓ | BS⇓ | ECE⇓ | AECE⇓ | OE ⇓ | MCE⇓ | BS⇓ |
| 1. Baseline model | 0.22 ± 0.03 | 0.19 ± 0.03 | **0.007** ± 0.004 | **0.73** ± 0.05 | 0.26 ± 0.02 | 0.52 ± 0.02 | 0.52 ± 0.02 | 0.19 ± 0.02 | 0.74 ± 0.03 | 0.40 ± 0.02 |
| 2. Paired Confidence Loss | 0.21 ± 0.02 | 0.18 ± 0.01 | 0.07 ± 0.03 | 0.91 ± 0.05 | 0.26 ± 0.02 | 0.46 ± 0.03 | 0.47 ± 0.02 | **0.15** ± 0.02 | **0.72** ± 0.01 | **0.38** ± 0.03 |
| 3. Probability Loss | 0.21 ± 0.03 | 0.18 ± 0.03 | 0.03 ± 0.02 | 0.90 ± 0.09 | 0.26 ± 0.01 | 0.55 ± 0.05 | 0.54 ± 0.06 | 0.22 ± 0.01 | 0.76 ± 0.04 | 0.40 ± 0.06 |
| 4. Confidence Weight | **0.19** ± 0.01 | **0.17** ± 0.01 | 0.02 ± 0.01 | 0.81 ± 0.09 | 0.23 ± 0.01 | **0.39** ± 0.01 | **0.40** ± 0.01 | 0.21 ± 0.03 | 0.82 ± 0.02 | 0.59 ± 0.02 |
| 5. AvUC Loss | 0.18 ± 0.02 | **0.13** ± 0.04 | **0.003** ± 0.001 | 0.96 ± 0.01 | **0.24** ± 0.01 | 0.50 ± 0.04 | 0.50 ± 0.04 | 0.21 ± 0.02 | 0.75 ± 0.01 | 0.43 ± 0.02 |
| 6. Soft ECE Loss | 0.20 ± 0.02 | 0.16 ± 0.02 | 0.03 ± 0.01 | 0.97 ± 0.01 | 0.25 ± 0.01 | 0.43 ± 0.03 | **0.42** ± 0.03 | **0.13** ± 0.00 | 0.73 ± 0.03 | 0.43 ± 0.03 |
| 7. MMCE Loss | 0.20 ± 0.01 | **0.16** ± 0.01 | 0.02 ± 0.00 | 0.95 ± 0.02 | 0.25 ± 0.02 | 0.47 ± 0.02 | 0.47 ± 0.03 | 0.15 ± 0.02 | 0.75 ± 0.01 | 0.40 ± 0.01 |

**Table 5**

All calibration metrics (Expected Calibration Error (ECE), Adaptive ECE (AECE), Overconfidence Error (OE), Maximum Calibration Error (MCE) and Brier Score(BS)) computed for the CRT models for epistemic (left) and aleatoric uncertainty (right). The lowest and optimal metric across strategies is highlighted in bold. All metrics should ideally move to zero as models become more calibrated as indicated by the ⇓.

| Method | CRT Epistemic | | | | | CRT Aleatoric | | | | |
|---|---|---|---|---|---|---|---|---|---|---|
| | ECE ⇓ | AECE⇓ | OE ⇓ | MCE ⇓ | BS⇓ | ECE⇓ | AECE⇓ | OE ⇓ | MCE⇓ | BS⇓ |
| 1. Baseline model | 0.28 | 0.22 | 0.10 | 0.92 | 0.29 | 0.33 | 0.36 | 0.17 | **0.70** | **0.29** |
| 2. Paired Confidence Loss | 0.32 | 0.27 | **0.05** | 0.6 | 0.28 | 0.35 | 0.30 | 0.20 | 0.95 | 0.33 |
| 3. Probability Loss | **0.25** | 0.23 | 0.08 | 0.95 | 0.30 | 0.34 | 0.31 | **0.14** | 0.95 | **0.29** |
| 4. Confidence Weight | 0.26 | **0.18** | 0.07 | **0.48** | **0.27** | 0.38 | 0.41 | **0.14** | 0.95 | 0.36 |
| 5. AvUC Loss | 0.34 | 0.29 | 0.09 | 0.95 | 0.29 | **0.30** | **0.11** | 0.15 | 0.70 | **0.29** |
| 6. Soft ECE Loss | 0.29 | 0.22 | 0.14 | 0.95 | 0.30 | **0.30** | **0.03** | 0.16 | 0.95 | 0.33 |
| 7. MMCE Loss | 0.28 | 0.22 | 0.11 | 0.95 | 0.30 | 0.40 | 0.39 | 0.18 | 0.78 | 0.31 |

**Table 6**

All calibration metrics (Expected Calibration Error (ECE), Adaptive ECE (AECE), Overconfidence Error (OE), Maximum Calibration Error (MCE) and Brier Score(BS)) computed for the CAD models for epistemic (left) and aleatoric uncertainty (right). The lowest and optimal metric across strategies is highlighted in bold All metrics should ideally move to zero as models become more calibrated as indicated by the ⇓.

| Method | CAD Epistemic | | | | | CAD Aleatoric | | | | |
|---|---|---|---|---|---|---|---|---|---|---|
| | ECE ⇓ | AECE⇓ | OE ⇓ | MCE ⇓ | BS⇓ | ECE⇓ | AECE⇓ | OE ⇓ | MCE⇓ | BS⇓ |
| 1. Baseline model | 0.46 | 0.61 | 0.17 | 0.76 | **0.21** | 0.51 | 0.58 | 0.30 | 0.83 | 0.43 |
| 2. Paired Confidence Loss | 0.48 | 0.59 | 0.12 | 0.87 | 0.24 | 0.49 | 0.63 | 0.17 | 0.83 | 0.21 |
| 3. Probability Loss | 0.48 | 0.60 | 0.21 | 0.72 | 0.23 | 0.51 | 0.58 | 0.29 | 0.79 | 0.40 |
| 4. Confidence Weight | **0.41** | **0.50** | 0.29 | **0.71** | **0.21** | 0.51 | 0.62 | 0.42 | **0.77** | **0.13** |
| 5. AvUC Loss | 0.46 | 0.56 | 0.13 | 0.78 | 0.23 | **0.47** | 0.58 | 0.19 | 0.79 | 0.3 |
| 6. Soft ECE Loss | 0.45 | 0.53 | **0.09** | 0.84 | 0.25 | 0.48 | 0.62 | **0.15** | 0.87 | 0.23 |
| 7. MMCE Loss | 0.45 | 0.55 | 0.15 | 0.95 | 0.25 | 0.48 | 0.58 | 0.19 | 0.83 | 0.34 |

in the segmentation model approximates the epistemic uncertainty of the segmentation model. However, the multiple segmentations generated using this approximation are passed as inputs into the VAE and classification model, in this way they can be used to approximate the aleatoric uncertainty of the VAE/classifier.

The epistemic uncertainty of the baseline and uncertainty-aware model was estimated using random sampling in the latent space of the VAE. Again, the original embedding together with 19 additional random samples were used for estimating epistemic uncertainty. Increasing the number of samples from the latent space did not have a statistically significant difference on the estimate but did adversely affect execution time, therefore just 20 samples were used for epistemic uncertainty estimation. For both types of uncertainty, the outputs of the Softmax layer were used to compute prediction confidence/uncertainty as a percentage of positive predictions out of the 20 samples. The values for all metrics for the epistemic and aleatoric uncertainty for both CRT response and CAD diagnosis models are presented in Tables 5 and 6 respectively, with the lowest and optimal metric highlighted in bold.

Interestingly, we see different outcomes for both CRT and CAD in the presence of epistemic and aleatoric uncertainty. The results indicate that the Confidence Weight model has a lower ECE than the baseline model for epistemic uncertainty, as can be seen in Table 5 but a similar outcome and consistency was not seen in the presence of aleatoric uncertainty for CRT. Similar to CRT, the CAD results in Table 6 highlight the same outcomes. Interestingly it may imply the modelling of aleatoric uncertainty may need further refinement. However, one may also argue that the ECE value might not be an optimal metric to utilise to assess calibration performance as our application is in a high risk setting, and therefore the OE measure could be a more appropriate metric. However, analysing the OE, a consistent outcome across both applications was not seen but did match the representation on the reliability diagram. One noticeable outcome for the CRT application was that the Confidence Weight model did seem to handle uncertainty with noticeably lower OE values.

### 5.4. Comparison of validation accuracy and calibration metric-based model selection

In this section we continue to analyse our uncertainty-aware training methods by investigating two different approaches for model selection. We use only the CAD diagnosis application for this analysis due to its larger training and test set sizes.

Most current research utilises the highest validation accuracy to identify the best/optimal performing model (up until this point we have used BACC). However, in our work we aim to provide more evidence of the optimal uncertainty-aware model by investigating if different optimal models would be obtained if we instead used lowest validation ECE as the criterion for model selection. We chose ECE as it is still the most common and widely utilised calibration measure, even with its weaknesses (Roelofs et al., 2022). We illustrate how the use of ECE and BACC as model selection criteria can affect the optimal performing model by indicating the test ECE and test BACC in Figs. 4 and 5 respectively. Here, the orange bars indicate the result when using validation ECE as the model selection criterion and the blue bars are the results when using validation BACC as the selection criterion.

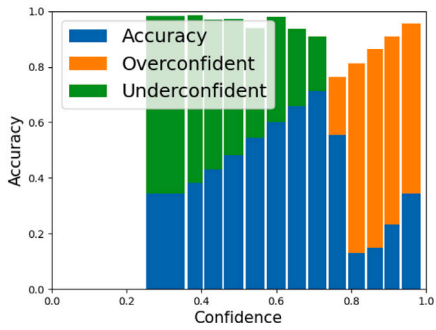

(a) Baseline

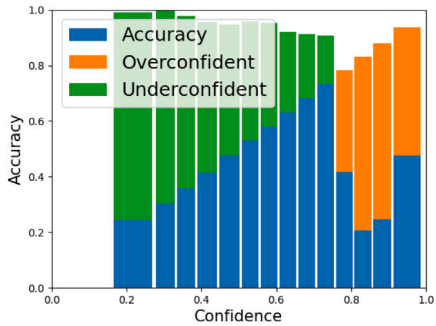

(b) Paired Confidence Loss

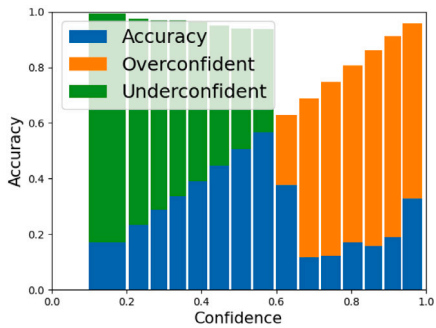

(c) Probability Loss

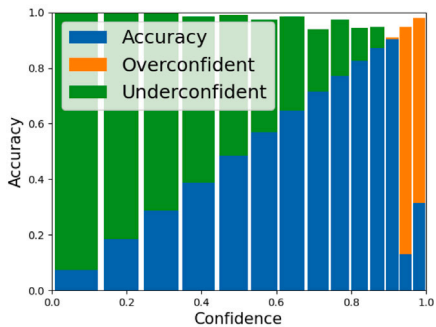

(d) Confidence Weight

**Fig. 2.** Reliability diagrams using adaptive binning to illustrate the performance of all strategies against the baseline model, indicating movement in samples and the relative accuracy and confidence relationship.

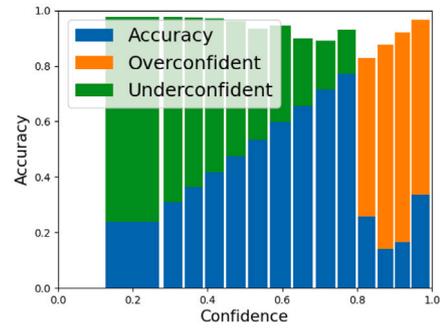

(e) Accuracy Versus Uncertainty Loss

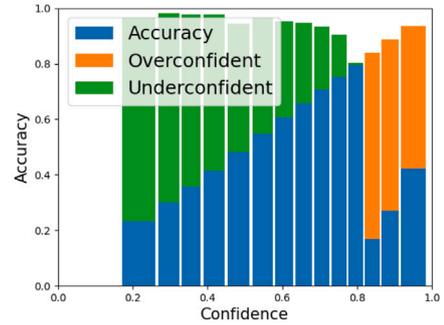

(f) Soft ECE Loss Function

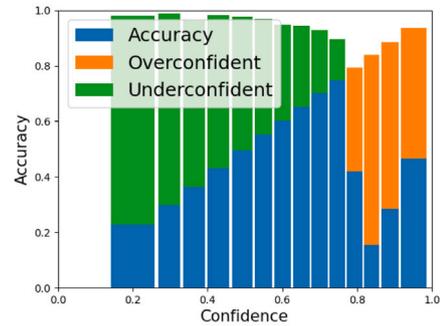

(g) MMCE Loss

**Fig. 3.** Reliability diagrams using adaptive binning to illustrate the performance of all strategies against the baseline model, indicating movement in samples and the relative accuracy and confidence relationship (Ding et al., 2020a).

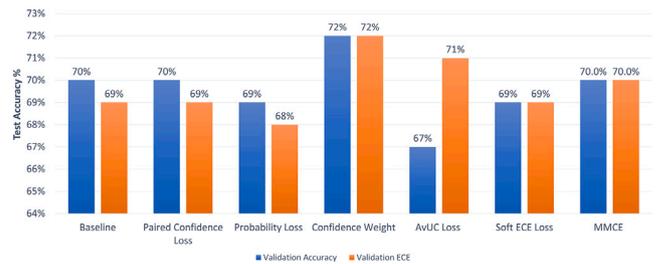

**Fig. 4.** Comparison of test BACC for the best-performing model within each uncertainty-aware method for different model selection criteria (validation BACC vs. validation ECE).

# 6. Discussion

In this paper we have proposed three novel uncertainty-aware training approaches, our Paired Confidence Loss from our preliminary investigation (Dawood et al., 2021), a Probability Loss function and a Confidence Weight term. Three comparative state-of-the-art approaches were also evaluated, Accuracy versus Uncertainty Loss, Soft ECE and MMCE Loss. All six strategies were evaluated for two clinically realistic CMR-based classification problems with the aim of finding a preferred uncertainty-aware strategy that can promote clinical trust in a decision support setting. Specifically, we want to reduce confident incorrect predictions and improve confidence in correct predictions. In our work we utilised both accuracy and calibration measures to identify the best performing model and also investigated different approaches for model selection, using the highest validation BACC versus the lowest validation ECE.

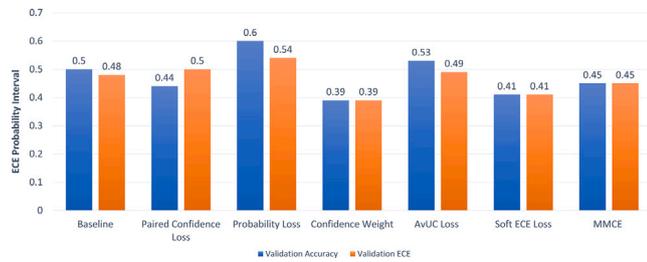

**Fig. 5.** Comparison of test ECE for the best-performing model within each uncertainty-aware method for different model selection criteria (validation BACC vs. validation ECE).

### 6.1. Model performance

Overall, according to the most commonly used calibration metric (ECE), our novel Confidence Weight strategy performed the best across both the CRT and CAD applications. However, for the CAD diagnosis model, the MCE for one of the bins in the Confidence Weight strategy indicated a high calibration error of 0.84, which may be attributed to the large deviation away from ideal calibration for lower confidence samples. However, considering that our goal for a high risk application is to identify and reduce overconfident wrong predictions, these low confidence bins might be less important. In our setting, after analysing our results, we argue that the overconfidence error may be a better measure to evaluate uncertainty-aware training methods, focusing as it does on overconfident wrong predictions. By this measure, the best-performing models for the CAD diagnosis task are the Paired Confidence Loss and the Soft ECE Loss.

However, our results highlight a fundamental difficulty with assessing model calibration using a single metric such as ECE. Specifically, our results tend to indicate that the calibration metrics do not completely agree. As an example, for the CAD diagnosis problem the 'best' model according to ECE actually increases the overconfidence error, maximum calibration error and Brier score. Likewise, on the same task the best-performing model according to Brier score is the Paired Confidence Loss, however this does not reduce ECE significantly. Analysis of the reliability diagram does allow us to explain some of the differences between ECE and Brier score, as Brier score is known to be insensitive to lower probabilities if fewer and infrequent samples lie within these bands (Ovadia et al., 2019). Analysing the overconfidence error, which we believe has the potential to be more useful for high risk applications, we see that the Confidence Weight model is no longer the best-performing model when analysed as a stand-alone calibration metric.

### 6.2. Model selection

Interestingly, we found that for the Soft ECE loss and the Confidence Weight strategies the optimal performing model was not affected by the model selection criterion. When analysing the baseline and other uncertainty-aware strategies, a surprising result can be observed: choosing the model based on validation BACC yielded better ECE values but the accuracies achieved were lower. However, some of these differences were relatively small and so require further investigation. We also note that the AvUC method had an optimal model when utilising the validation ECE but had poorer performance if the best validation BACC was utilised.

Our analysis suggests that the choice of model selection criterion may be important for uncertainty-aware training methods, a point that we do not believe has been highlighted before in the literature. However, it appears that there is no single correct model selection measure that will consistently achieve good model calibration outcomes.

Overall, we argue that the best approach may be to look at a range of model selection metrics and choose the model that maximises both accuracy and calibration, with the calibration metric(s) being chosen to suit the context of the intended application.

### 6.3. Limitations and future work

In our work we made use of Softmax probabilities, which are widely utilised and accepted but are known to be less calibrated estimates of uncertainty (Gupta et al., 2020). Additionally, our VAE architecture using multiple time-based image stacks may have prevented robust estimates of uncertainty and limited calibration performance. In future work we will aim to incorporate alternative direct methods of uncertainty estimation during training of DL models, to reduce overestimation and underestimation of confidence, which is known to be an ongoing research problem within the field of uncertainty estimation and model calibration.

Future work will also focus on more extensive investigation and analysis of uncertainty-aware training methods for a wider range of clinical problems. We will investigate the development of alternate calibration metrics which are more tuned to specific (clinical) contexts and/or are less biased and more applicable to the healthcare setting. Furthermore we will investigate alternate architectures for quantifying uncertainty in a robust manner as well as alternate strategies for improving calibration such as focal loss (Kumar and Sarawagi, 2019). Additionally, we plan to investigate the impact of label smoothing (Carse et al., 2022) on our uncertainty-aware approaches. In this paper we chose to focus on uncertainty-aware training methods, rather than approaches that alter the training labels, but we note that label smoothing approaches could be combined with any uncertainty-aware training method, and the interaction of these two approaches should be thoroughly investigated. We will also investigate the possibility of using other calibration metrics, such as overconfidence error, for model selection, rather than BACC and ECE as we have investigated in this paper. In addition, we believe that it is important to evaluate the impact of AI on clinical workflows in a decision support setting, and the importance of model calibration on this impact. Future work will also focus on this area.

## 7. Conclusion

In summary, we have investigated a range of different calibration metrics to assess our uncertainty-aware training methods. In terms of the most commonly used calibration metric (ECE), the Confidence Weight approach resulted in the best-calibrated models. However, we highlighted that the choice of best model would vary depending on the metric used. We have argued that overconfidence error might be the most appropriate metric for high risk medical applications, and in terms of overconfidence error the best-performing models were the Paired Confidence Loss and the Soft ECE Loss.

Overall our analysis indicated that the goal of trying to improve deep learning model calibration for cardiac MR applications was achieved but only in terms of some calibration metrics. The results further highlighted the potential weakness of current measures and indicated the need to continue to investigate and identify robust metrics for high risk healthcare applications rather than simply using ECE and BACC (Gupta et al., 2020), bearing in mind that the most relevant metrics may not be the same for different applications. Further research into uncertainty-aware training for optimising different (combinations of) metrics is also recommended.

## Declaration of competing interest

The authors declare the following financial interests/personal relationships which may be considered as potential competing interests: Tareen Dawood reports financial support was provided by NIHR Biomedical Research Centre at Guy's and Saint Thomas' NHS Foundation Trust and King's College. Esther Puyol reports financial support was provided by Wellcome Trust.

## Data statement

The UKBB datasets presented in this study are publicly available and can be found in online repositories under approved research projects from https://www.ukbiobank.ac.uk/. The GSTFT dataset cannot be made publicly available due to restricted access under hospital ethics and because informed consent from participants did not cover public deposition of data.

## Acknowledgements

This work was supported by the Kings DRIVE Health CDT for Data-Driven Health and further funded/supported by the National Institute for Health Research (NIHR) Biomedical Research Centre at Guy's and St Thomas' NHS Foundation Trust and King's College London, United Kingdom. Additionally this research was funded in whole, or in part, by the Wellcome Trust, United Kingdom WT203148/Z/16/Z. For the purpose of open access, the author has applied a CC BY public copyright licence to any author accepted manuscript version arising from this submission. The work was also supported by the EPSRC, United Kingdom through the SmartHeart Programme Grant (EP/P001009/1). This research has been conducted using the UK Biobank Resource under Application Number 17806. The views expressed in this paper are those of the authors and not necessarily those of the NHS, EPSRC, the NIHR or the Department of Health and Social Care.